\documentclass[cameraready]{Interspeech}

\title{Speech Enhancement Based on Drifting Models}

\usepackage{amsmath}
\usepackage{xcolor}
\usepackage{booktabs}
\usepackage{cite}
\usepackage{graphicx}

\author[affiliation={1}]{Liang}{Xu}
\author[affiliation={2}]{Diego}{Caviedes-Nozal}
\author[affiliation={1}]{W. Bastiaan}{Kleijn}
\author[affiliation={1,3}]{\\Longfei Felix}{Yan}
\author[affiliation={2}]{Rasmus Kongsgaard}{Olsson}

\address{
    $^1$ Victoria University of Wellington, $^3$ Lincoln University, New Zealand \\
    $^2$ GN Advanced Science, Denmark
}

\email{
{liang.xu,bastiaan.kleijn}@vuw.ac.nz\\
{dcnozal,rkolsson}@gn.com; felix.yan@lincoln.ac.nz
}

\usepackage{comment}
\usepackage{amssymb}

\usepackage{amsfonts}

\begin{document}
\maketitle
\begin{abstract}
    We propose Speech Enhancement based on Drifting Models (DriftSE), a novel generative framework that formulates denoising as an equilibrium problem. Rather than relying on iterative sampling, DriftSE natively achieves one-step inference by evolving the pushforward distribution of a mapping function to directly match the clean speech distribution. This evolution is driven by a Drifting Field, a learned correction vector that guides samples toward the high-density regions of the clean distribution, which naturally facilitates training on unpaired data by matching distributions rather than individual paired samples. We investigate the framework under two formulations: a direct mapping from the noisy observation, and a stochastic conditional generative model from a Gaussian prior. Experiments on the VoiceBank-DEMAND benchmark demonstrate that DriftSE achieves high-fidelity enhancement in a single step, outperforming multi-step diffusion baselines and establishing a new paradigm for speech enhancement.
\end{abstract}

\section{Introduction}
\label{sec:intro}

The field of Speech Enhancement (SE) has evolved significantly over recent decades, progressing from classical statistical signal processing techniques like Wiener filtering~\cite{meyer1997multi, chua2024effective} to modern deep learning. Discriminative models, such as RNNs~\cite{Weninger2015LVA}, LSTMs~\cite{tan18_interspeech}, and complex spectral mapping~\cite{hu2020dccrn}, effectively suppress noise but often yield spectral oversmoothing and robotic artifacts due to regression-based objectives. Generative Adversarial Networks (GANs)~\cite{pascual2017segan, fu2019metricgan, su2021hifi} improve perceptual quality but suffer from training instability and mode collapse. Recently, Score-based Diffusion Models~\cite{richter2023speech} have established state-of-the-art performance by modeling the gradient of the log-density of the clean speech distribution. These models define a forward process that gradually degrades data into noise, and a reverse process for generation. The reverse dynamics can be formulated as either a Stochastic Differential Equation (SDE)~\cite{song2021scorebased} or a deterministic Probability Flow ODE (PF-ODE) sharing the same marginal probability densities. However, their inference is inherently iterative. Numerically integrating these highly curved reverse-time trajectories requires 10–100 discretization steps, resulting in a high Number of Function Evaluations (NFE) that imposes a critical latency bottleneck for real-time applications.

To address the computational inefficiency of diffusion models, recent research broadly falls into two lines of work: trajectory compression and trajectory linearization. Compression approaches accelerate sampling by reducing the number of steps. For instance, hybrid approaches~\cite{lemercier2023storm, trachu24_interspeech} combine predictive models with a small number of diffusion refinement steps, while diffusion-GAN hybrids~\cite{han25b_interspeech} further reduce steps via adversarial training. Similarly, distillation-based one-step generators such as Consistency Models~\cite{song2023consistency,kim2024consistency,xu2025rosecd, nishigori2025schrodinger}  enforce self-consistency along the PF-ODE to distill a multi-step sampler into a single-step mapping. 

In parallel, Flow Matching~\cite{lipman2023flow, wang2025flowse} techniques seek to linearize the generative trajectory. Rectified Flow~\cite{rectflow2023} explicitly straightens the transport path to minimize the curvature of the ODE. MeanFlow~\cite{geng2025meanflow, geng2025improved} learns a continuous mean velocity field to model probability paths. However, these methods remain fundamentally trajectory-based. They rely on continuous transport dynamics that must be discretized at inference, and accurately approximating these paths with only a few steps remains challenging.

Recently, Drifting Models~\cite{deng2026generative} were proposed as a powerful new paradigm that reformulates generation as a distributional equilibrium problem. By mapping high-dimensional data into a semantic latent space during training, the framework learns a kernelized drifting field where generated samples are simultaneously attracted to the true data distribution and repelled by the evolving model distribution. Minimizing this drift directly aligns the generator's pushforward distribution with the target data. Operating natively in a single step, this latent equilibrium approach has achieved state-of-the-art results in large-scale image generation (FID 1.54 on ImageNet).

Our contribution is the introduction of Speech Enhancement based on Drifting Models (DriftSE), a novel generative framework for one-step denoising. DriftSE formulates enhancement as learning a direct projection onto the clean speech manifold without imposing predefined trajectory constraints. As a purely generative model, it naturally supports learning from completely unpaired data. We adapt DriftSE for two distinct enhancement paradigms: a direct mapping that pushes the noisy speech distribution toward the clean speech distribution, and a stochastic conditional generative approach that generates clean speech from a Gaussian noise prior. To construct a perceptually meaningful drifting field, we project the audio into a semantic latent space using a pre-trained encoder. Aligning these latent distributions ensures high quality SE.

Extensive experiments on the VoiceBank-DEMAND dataset demonstrate that DriftSE achieves competitive perceptual quality with single-step inference. Specifically, the direct mapping variant achieves PESQ 3.15 and SI-SDR 16.1~dB, while the conditional variant achieves SCOREQ 4.33. Evaluation on real-world recordings from the DNS Challenge 2020 blind test set demonstrates state-of-the-art generalization performance. Furthermore, we demonstrate that DriftSE can perform speech enhancement in fully unpaired cross-dataset and cross-gender settings, highlighting the flexibility of the proposed generative formulation.

\section{Drifting Models}
\label{sec:drifting_models}

We briefly review Drifting Models~\cite{deng2026generative}, which formulate generative modeling as the training-time evolution of a pushforward distribution. 

\subsection{Pushforward and Equilibrium}
\label{ssec:pushforward}

Given a simple source distribution $p_{\epsilon}$ (e.g., standard Gaussian noise $\mathcal{N}(\mathbf{0}, \mathbf{I})$), the drift approach takes a sample $\epsilon \sim p_{\epsilon}$ with $\epsilon \in \mathbb{R}^{d}$, and maps it through a parameterized function $f_\theta: \mathbb{R}^{d} \rightarrow \mathbb{R}^{d}$ in a single step to produce a target variable $\mathbf{x} \in \mathbb{R}^{d}$
\begin{equation}\label{eq:target}
    \mathbf{x} = f_{\theta}(\epsilon).
\end{equation}
This defines the pushforward distribution $q_\theta = (f_\theta)_\# p_{\epsilon}$, meaning that sampling $\epsilon \sim p_{\epsilon}$ and applying $f_\theta$ yields samples distributed as $q_\theta$. For conditional generation, ~\eqref{eq:target} extends to $\mathbf{x} = f_{\theta}(\epsilon, \mathbf{c})$, where $\mathbf{c}$ is a condition (e.g., a class label or noisy speech).

To drive $q_{\theta}$ toward the target data distribution $p_{\text{data}}$, the framework introduces a Drifting Field $\mathbf{V}_{p,q}:\mathbb{R}^d\rightarrow\mathbb{R}^d$ that acts as a correction vector at each generated sample of $q_\theta$, pointing in the direction that reduces the discrepancy between $q_{\theta}$ and $p_{\text{data}}$. Concretely, a drifting target is defined as
\begin{equation}
    \mathbf{x}_{\text{target}} \leftarrow \mathbf{x} + \mathbf{V}_{p,q}(\mathbf{x}),
    \label{eq:update_rule}
\end{equation}
and the generator is trained to map $\epsilon$ to $\mathbf{x}_{\text{target}}$.
As training progresses, the pushforward distribution $q_{\theta}$ evolves until it reaches a state of distributional equilibrium where the drift vanishes
\begin{equation}
    q_{\theta} = p_{\text{data}}
    \quad \Longrightarrow \quad
    \mathbf{V}_{p,q}(\mathbf{x}) = \mathbf{0},\ \forall \mathbf{x}.
    \label{eq:equilibrium}
\end{equation}

\subsection{Designing the Drifting Field}
\label{ssec:designing_field}

Inspired by mean-shift theory~\cite{meanshift1995}, the total drift $\mathbf{V}_{p,q}(\mathbf{x})$ decomposes into two opposing forces
\begin{equation}
    \mathbf{V}_{p,q}(\mathbf{x}) = \mathbf{V}^+_p(\mathbf{x}) - \mathbf{V}^-_q(\mathbf{x}),
    \label{eq:drift_decomposition}
\end{equation}
where $\mathbf{V}^+_{p}$ attracts samples toward the data distribution $p_{\text{data}}$, and $\mathbf{V}^-_{q}$ repels samples away from high-density regions of the current model distribution $q_{\theta}$.
Both terms take the form of a kernel-weighted mean shift
\begin{align}
    \mathbf{V}^+_p(\mathbf{x}) &= \frac{1}{Z_p(\mathbf{x})} \mathbb{E}_{\mathbf{y}^+ \sim p} \left[ k(\mathbf{x}, \mathbf{y}^+) (\mathbf{y}^+ - \mathbf{x}) \right], \\
    \mathbf{V}^-_q(\mathbf{x}) &= \frac{1}{Z_q(\mathbf{x})} \mathbb{E}_{\mathbf{y}^- \sim q} \left[ k(\mathbf{x}, \mathbf{y}^-) (\mathbf{y}^- - \mathbf{x}) \right],
\end{align}
with similarity kernel $k(\mathbf{x},\mathbf{y})$ and normalizers
$Z_p(\mathbf{x})=\mathbb{E}_{\mathbf{y}^{+}\sim p}[k(\mathbf{x},\mathbf{y}^{+})]$ and
$Z_q(\mathbf{x})=\mathbb{E}_{\mathbf{y}^{-}\sim q}[k(\mathbf{x},\mathbf{y}^{-})]$.

In practice, these expectations are approximated with mini-batch averages, drawing positives $\mathbf{y}^+$ from $p_{\text{data}}$ and negatives $\mathbf{y}^-$ from the current model distribution $q_{\theta}$. By substituting the normalizers and combining the forces into a joint expectation over $p$ and $q$, the self-referential $-\mathbf{x}$ terms elegantly cancel out. This yields the exact unified formulation
\begin{equation}
\label{eq:joint_drift_field}
    \mathbf{V}_{p,q}(\mathbf{x}) = \frac{1}{Z_p Z_q} \mathbb{E}_{p,q} \left[ k(\mathbf{x}, \mathbf{y}^+) k(\mathbf{x}, \mathbf{y}^-) (\mathbf{y}^+ - \mathbf{y}^-) \right].
\end{equation}

To measure the similarity between a generated sample $\mathbf{x}$ and any reference feature $\mathbf{y} \in \{\mathbf{y}^+, \mathbf{y}^-\}$, an exponential similarity kernel with temperature $\tau$ is employed
\begin{equation}
\label{eq:kernel_exp}
    k_\tau(\mathbf{x}, \mathbf{y}) = \exp\left(-\frac{\left\Vert \mathbf{x} - \mathbf{y} \right\Vert_2}{\tau}\right),
\end{equation}
where $\tau$ controls the interaction bandwidth.

\subsection{Training Objective}
\label{ssec:training_objective}
In practice, the drifting field is computed in a latent space using a pretrained extractor $\phi(\cdot)$. To optimize the generator $f_\theta$, its outputs $\mathbf{x} = f_\theta(\epsilon)$ are driven along the field $\mathbf{V}$ via the objective
\begin{equation}
    \mathcal{L}_{\text{drift}} = \mathbb{E}_{\epsilon} \left[ \Big\Vert \phi(\mathbf{x}) - \text{sg} \left( \phi(\mathbf{x}) + \mathbf{V}\big(\phi(\mathbf{x})\big) \right) \Big\Vert_2^2 \right],
    \label{eq:loss_drift}
\end{equation}
where $\text{sg}(\cdot)$ is the stop-gradient operator. Regressing toward this fixed target minimizes the magnitude of $\mathbf{V}$, progressively transporting the pushforward distribution $q_{\theta}$ toward the target data distribution $p_{\text{data}}$.

\begin{figure*}[t]
  \centering
  \includegraphics[width=\textwidth]{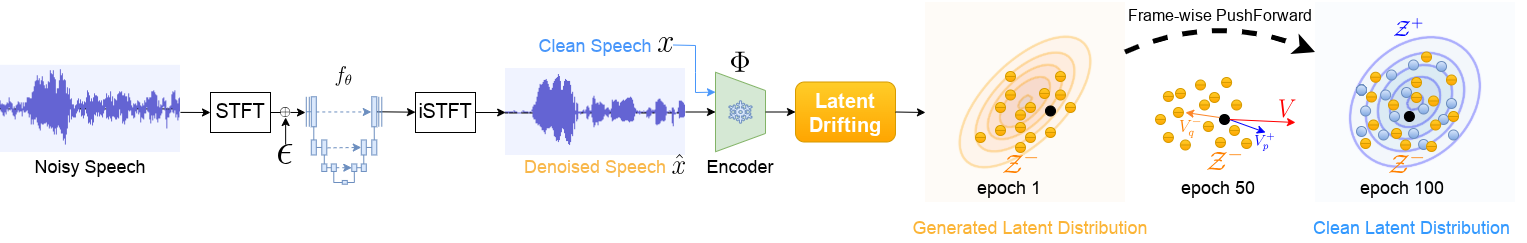}
  \caption{Overview of the DriftSE framework (illustrating the Direct Mapping formulation).}
  \label{fig:drifting_paradigms}
  \vspace{-6pt}
\end{figure*}

\section{Speech Enhancement via Latent Drifting}
\label{sec:drifting_se}
We propose DriftSE, which formulates speech enhancement as an equilibrium problem (Fig.~\ref{fig:drifting_paradigms}). By evolving the mapping function's pushforward distribution to match the clean speech distribution, DriftSE achieves native one-step denoising (1 NFE).

\subsection{Two Enhancement Paradigms}

Let $\mathbf{y} \in \mathbb{C}^{F \times T}$ denote the complex spectrogram of the noisy speech, with $F$ frequency bins and $T$ time frames, and let $\mathbf{x} \in \mathbb{C}^{F \times T}$ be the clean speech target. To produce the enhanced speech $\hat{\mathbf{x}}$, we investigate two distinct formulations for the mapping function $f_\theta$:

\textbf{Direct Mapping:} Defined as $\hat{\mathbf{x}} = f_\theta(\mathbf{y} + \sigma \boldsymbol{\epsilon})$, where $\boldsymbol{\epsilon} \sim \mathcal{N}(\mathbf{0}, \mathbf{I})$ and $\sigma$ controls the noise injection strength. During training, when $\sigma = 0$, this acts as a strictly deterministic mapping $\hat{\mathbf{x}} = f_\theta(\mathbf{y})$. When $\sigma > 0$, the injected noise smooths the acoustic distribution, aiding the estimation of the drifting field.

\textbf{Conditional Generator:} Defined as $\hat{\mathbf{x}} = f_\theta(\boldsymbol{\epsilon}, \mathbf{y})$, where the network maps a standard Gaussian noise prior $\boldsymbol{\epsilon}$ to the target distribution conditioned on $\mathbf{y}$. 

\subsection{Speech Latent Encoder}

Applying the drifting field in the time-frequency domain is suboptimal, as Euclidean distances on raw spectrograms are dominated by high-amplitude harmonics, neglecting the low-energy transients critical for phonetic intelligibility. Following \cite{deng2026generative}, we instead compute drift in a semantic latent space and apply multi-layer latent supervision, which provides a richer and more stable training signal. For speech enhancement, we selected self-supervised speech models. In particular, we employed HuBERT, WavLM and DistilHuBERT~\cite{hsu2021hubert, chen2022wavlm, chang2022distilhubert}, which exhibit a well-documented layer hierarchy: shallow layers capture low-level acoustic structure, while deeper layers encode phonetic and semantic content. We therefore define a frozen self-supervised learning (SSL) encoder $\Phi: \mathbb{R}^{L} \rightarrow \mathbb{R}^{T' \times D}$ that maps a waveform of length $L$ to $T'$ frame-level latent features of dimension $D$. To capture the hierarchical speech structures, the drifting field is computed and aggregated across a selected set of layers $\mathcal{S}$.

\subsection{Frame-Wise Latent Drifting and Inference}
\label{ssec:drifting_and_inference}

\textbf{Latent Drifting:} As detailed in Fig.~\ref{fig:drifting_paradigms}, we dynamically construct a positive set of samples $\mathcal{Z}^+$ from clean reference frames $\Phi(\mathbf{x})$ and a negative set of samples $\mathcal{Z}^-$ from the current batch of generated frames $\Phi(\hat{\mathbf{x}})$. For any generated frame $\mathbf{z}_i \in \mathcal{Z}^-$, the frame-wise Drifting Field $\mathbf{V}(\mathbf{z}_i)$ is computed by instantiating ~\eqref{eq:joint_drift_field} in the speech latent space, using the multi-temperature kernel ${k}_\tau$ from ~\eqref{eq:kernel_exp}. The resulting field combines an attraction force pulling $\mathbf{z}_i$ toward the clean distribution $\mathcal{Z}^+$ and a repulsion force pushing it away from the current generated distribution $\mathcal{Z}^-$, driving $f_\theta$ toward equilibrium.

\textbf{Training Objective:} To capture hierarchical speech structures, the base drifting loss from ~\eqref{eq:loss_drift} is computed and aggregated across multiple layers $l \in \mathcal{S}$ of the latent encoder.

\textbf{Inference:} At inference time, the direct mapping approach uses $\sigma = 0$ for deterministic denoising, while the conditional generator draws a fresh $\boldsymbol{\epsilon}$ to generate diverse enhanced outputs. 

\textbf{Overview of DriftSE:} Fig.~\ref{fig:drifting_paradigms} illustrates the method. The mapping function $f_\theta$ processes the noisy speech spectrogram alongside injected Gaussian noise $\boldsymbol{\epsilon}$, which acts as a distribution smoother, to produce a denoised spectrogram in a single step. After iSTFT, both the enhanced waveform $\hat{\mathbf{x}}$ and the clean reference $\mathbf{x}$ are projected into a frame-wise latent space via a frozen encoder $\Phi$. For frame-wise latent drifting at each training iteration, we dynamically construct a mini-batch positive set $\textcolor{blue}{\mathcal{Z}^+}$ sampled from the clean feature frames $\Phi(\mathbf{x})$, and a mini-batch negative set $\textcolor{orange}{\mathcal{Z}^-}$ sampled from the mapped feature frames $\Phi(\hat{\mathbf{x}})$. The total Drifting Field $\textcolor{red}{\mathbf{V}}$ for a mapped frame is composed of an attraction force $\textcolor{blue}{\mathbf{V}_p^+}$ that pulls it toward high-density regions of the empirical target distribution $\textcolor{blue}{\mathcal{Z}^+}$, and a repulsion force $\textcolor{orange}{\mathbf{V}_q^-}$ that pushes it away from its neighbors in the current model distribution $\textcolor{orange}{\mathcal{Z}^-}$. As training progresses, the mapping function minimizes $\textcolor{red}{\mathbf{V}}$, dynamically evolving the pushforward distribution until it matches the clean distribution at equilibrium.

\textbf{Unpaired Learning Capability:} DriftSE natively supports fully unpaired learning. Although we utilize paired audio to maximize reconstruction fidelity as is standard practice in generative SE, our formulation remains fundamentally distribution-based. The internal drift relies strictly on latent similarity across all frames within the current batch, rather than enforcing rigid frame-to-frame regression.
\section{Experiments}
\label{sec:experiments}
In this section, we evaluate DriftSE against state-of-the-art iterative and one-step baselines, and perform ablation studies to analyze the contribution of each design choice.

\subsection{Experimental Setup}

\textbf{Datasets:} We train on clean speech from the VoiceBank corpus~\cite{botinhao2016investigating} and noise recordings from the DEMAND dataset~\cite{thiemann2013diverse}. During training, we employ dynamic mixing where 10,802 clean utterances are mixed on-the-fly with 18 distinct noise types~\cite{richter2023speech}. To ensure robust generalization and prevent overfitting to specific acoustic conditions, Signal-to-Noise Ratios (SNRs) are sampled randomly from $\{0, 5, 10, 15\}$~dB.

For evaluation, we utilize the standard pre-mixed VB-DMD test set (824 utterances) to ensure fair benchmarking. To assess real-world generalization, we further evaluate on the DNS Challenge 2020 blind test set~\cite{reddy2020interspeech}, which contains 300 real-world noisy recordings without clean references.

\textbf{Evaluation:} Following previous studies~\cite{richter2023speech, xu2025rosecd}, we report pairwise metrics including PESQ~\cite{rix2001perceptual}, ESTOI~\cite{jensen2016algorithm} and SI-SDR~\cite{le2019sdr}. To assess perceptual quality without clean reference, we also report non-intrusive metrics: SCOREQ~\cite{ragano2024scoreq}, DNSMOS~\cite{reddy2021dnsmos, reddy2022dnsmos}, and WV-MOS~\cite{andreev2023hifipp}.

\textbf{Implementation Details:} We use NCSN++V2~\cite{richter2023speech} without time embedding, with 16~kHz audio processed by STFT using a 510-point Hann window and hop length 128, followed by spectral compression~\cite{richter2023speech}. For the mapping DriftSE variant, we empirically sample the noise level $\sigma$ from a truncated log-normal distribution, i.e., $\log \sigma \sim \mathcal{N}(-3.0, 1.2)$, truncated to $\sigma \in [0.01, 0.3]$. For the conditional generative variant, the STFT spectrogram of $\mathbf{y}$ is provided as a conditioning embedding into the generator at each resolution level. For the SSL latent encoder $\Phi$, we utilize the pre-trained HuBERT-Large, WavLM-Large, and DistilHuBERT checkpoints. The extracted latent frames have a 20ms hop size and 25ms receptive field. We aggregate features from layers $\mathcal{S} = \{6, 12, 24\}$ for WavLM-Large and HuBERT-Large, and layers $\mathcal{S} = \{0, 1, 2\}$ for DistilHuBERT, with all layers equally weighted. We use a multi-temperature exponential kernel (~\eqref{eq:kernel_exp}) with temperatures $\tau \in \{0.1, 0.5, 1.0\}$. The model was trained for 100 epochs on an A6000 GPU using AdamW with a batch size of 16, a learning rate of $5 \times 10^{-4}$, and a weight decay of $0.01$.

\subsection{Results}

\textbf{In-domain Evaluation:} As shown in Table~\ref{tab:performance}, DriftSE (DistilHuBERT, $\sigma{=}0$) achieves high-fidelity enhancement and outperforms the 30-step SGMSE+~\cite{richter2023speech} and the one-step MeanFlowSE~\cite{li2026meanflowse}, reaching PESQ~3.15 and confirming that latent drifting effectively maps noisy observations onto the distribution support of the clean speech. While distillation methods such as ROSE-CD~\cite{xu2025rosecd} and SBCTM~\cite{nishigori2025schrodinger} report higher PESQ, they utilize auxiliary losses. When we incorporate the same losses, DriftSE~\footnote[1$^\dagger$]{Model jointly trained with auxiliary PESQ and SI-SDR losses.} attains competitive performance.

\textbf{Generalization Evaluation:} We evaluate the generalization capability of DriftSE on the DNS Challenge 2020 blind test set in Table~\ref{tab:dns_full}. We achieve state-of-the-art WV-MOS~2.65 and SCOREQ~2.97, outperforming other baselines while delivering highly competitive perceptual scores (DNSMOS SIG, BAK, and OVRL). This confirms that the drifting equilibrium learns a highly generalizable distributional projection.

\begin{figure*}[!t]
  \centering
  \includegraphics[width=\textwidth]{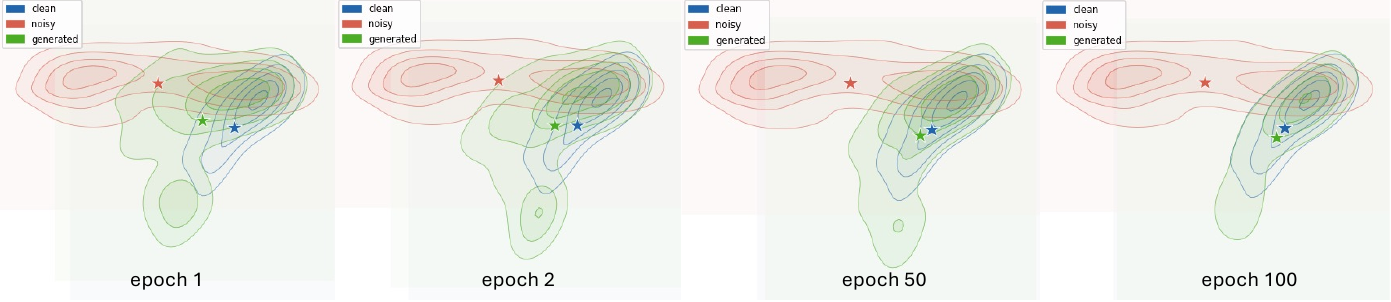}
    \caption{Evolution of frame-level distributions in the DistilHuBERT semantic space for a fixed test utterance. Each panel displays 2D density contours (PCA projection) derived from all frames across different training epochs. Stars denote the corresponding centroids, which represent the mean of all projected frames. As training progresses, the \textcolor{green!60!black}{generated distribution} shifts from the \textcolor{red!60}{noisy distribution} toward the \textcolor{blue!80}{clean distribution}.}
  \label{fig:pca_evolution}
  \vspace{-6pt}
\end{figure*}

\subsection{Ablations and Analysis}

\textbf{Impact of the Latent Encoder:} We evaluate the impact of different encoders and layer selections on latent drifting. Using only the deepest semantic layer (WavLM, Layer 24) degrades performance, suggesting that highly abstract features miss fine acoustic details. With multi-layer drifting, DistilHuBERT (768-d) is competitive with both HuBERT and WavLM (1024-d), achieving the best SI-SDR while maintaining similar perceptual quality. Therefore, we use DistilHuBERT as the default encoder in subsequent experiments.

\textbf{Conditional Drifting Models (DriftSE$^*$\footnote[2]{Conditional drifting model.}):} As shown in Table~\ref{tab:performance}, DriftSE$^*$ successfully reduces generative ambiguity, delivering superior reference-free perceptual metrics (DNSMOS 3.64, SCOREQ 4.33) while maintaining competitive pairwise fidelity. This demonstrates that incorporating a stochastic prior enables the generator's pushforward distribution to better capture the inherent variance of the clean speech distribution, leading to more natural generation.

\textbf{Effect of Noise Injection:} For the direct mapping variant, omitting the noise prior ($\sigma=0$) during training enforces a deterministic mapping with higher fidelity (PESQ 3.15, SI-SDR 16.10~dB), whereas injecting Gaussian noise smooths the acoustic distribution to improve reference-free perceptual quality (SCOREQ from 4.08 to 4.15). This distributional smoothing trades marginal waveform precision for more natural generation, providing a promising direction for adapting to narrow or shifted target distributions in future work. 

\textbf{Unpaired Learning:} To evaluate DriftSE in a fully unpaired setting, we train the model by sampling noisy and clean speech independently. We investigate two distinct training scenarios: a cross-dataset setup and a cross-gender setup.

In the cross-dataset setup for DriftSE (Unpaired, map to DNS), mapping noisy VoiceBank speech to clean DNS Challenge 2020 training targets achieves strong reference-free quality (DNSMOS 3.61, SCOREQ 3.92). Although pairwise fidelity predictably drops (PESQ 2.00) without noisy-clean pairing, the non-intrusive scores indicate successful distribution matching.

In the more challenging cross-gender setup for DriftSE (Unpaired, map to VB-Female), mapping mixed-gender noisy speech in VoiceBank to a restricted female-only target inherently forces the model to alter male speaker characteristics. Because this is a much harder generative task, perceptual scores naturally decrease (DNSMOS 3.40, SCOREQ 3.72). Consequently, reference-based metrics are omitted here as no ground-truth clean female references exist for the noisy male inputs.


\begin{table}[t]
\centering
\caption{Comparison on VB-DMD. NFE: Number of Function Evaluations. \textbf{Bold} indicates the best performing metric within each respective group.}
\label{tab:performance}
\resizebox{\linewidth}{!}{
\begin{tabular}{l c c c c c c}
    \toprule
    \textbf{Method} & \textbf{NFE} & \textbf{PESQ} & \textbf{SI-SDR} & \textbf{ESTOI} & \textbf{DNSMOS} & \textbf{SCOREQ} \\
    \midrule    
    MetricGAN+~\cite{fu2021metricganplus}                  & 1  & 3.13 & 8.50 & 0.83 & 3.22 & 3.82 \\
    UNIVERSE++~\cite{scheibler2024universeplusplus}     & 8  & 2.91 & 18.00 & 0.85 & 3.45 & \textbf{4.35} \\
    SGMSE+~\cite{richter2023speech}                     & 30 & 2.90 & 16.90 & 0.85 & 3.48 & 3.98 \\
    ROSE-CD~\cite{xu2025rosecd}                         & 1  & 3.49 & 17.80 & {0.87} & 3.49 & 4.23 \\
    SBCTM~\cite{nishigori2025schrodinger}               & 1  & \textbf{3.56} & 12.70 & {0.87} & 3.55 & \textbf{4.35} \\
    MeanFlowSE~\cite{li2026meanflowse}                  & 1  & 2.81 & \textbf{19.97} & \textbf{0.88} & \textbf{3.58} & 4.25 \\
    \midrule
    {DriftSE (WavLM, L24)} & 1 & 2.90& 12.60& 0.84& 3.36& 3.93\\    
    {DriftSE (WavLM)} & 1 & 3.03& 14.00& 0.85& \textbf{3.54}& \textbf{4.17}\\    
    {DriftSE (HuBERT)} & 1 & 2.94& 12.50& 0.84& 3.49& 4.14\\
    {DriftSE (DistilHuBERT)} &  1 & 3.00& 15.60& 0.85& 3.48& 4.15\\   
    {DriftSE (DistilHuBERT, $\sigma=0$)} & 1 & \textbf{3.15}& \textbf{16.10}& \textbf{0.86}& 3.47& 4.08\\    
    \midrule    
    {DriftSE$^*$ (DistilHuBERT)} & 1 & 2.99& 17.98& 0.86& \textbf{3.64}& \textbf{4.33}\\
    {DriftSE$^\dagger$ (DistilHuBERT)} & 1 & \textbf{3.45}& \textbf{20.60}& \textbf{0.87}& 3.49& 4.11\\
    \midrule        
    {{DriftSE (Unpaired, map to DNS)}} & 1 & 2.00& 6.60& 0.74& \textbf{3.61} & \textbf{3.92}\\
    {{DriftSE (Unpaired, map to VB-Female)}} & 1 & -& -& -& 3.40& 3.72\\    
    \bottomrule
\end{tabular}
}
\end{table}

\begin{table}[t]
\centering
\caption{Real-world recordings evaluation on DNS Challenge 2020 Blind Test Set. \textbf{Bold} indicates the best performing metric within each respective group.}
\label{tab:dns_full}
\resizebox{\linewidth}{!}{
\begin{tabular}{l c c c c c c}
    \toprule
    \textbf{Method} & \textbf{NFE} & \textbf{WV-MOS} & \textbf{SCOREQ} & \textbf{SIG} & \textbf{BAK} & \textbf{OVRL} \\
    \midrule
    MetricGAN+~\cite{fu2021metricganplus}                & 1  & 1.23 & 2.08 & 3.28 & 3.45 & 2.70 \\
    UNIVERSE++~\cite{scheibler2024universeplusplus}   & 8  & 1.99 & 2.27 & 3.45 & 3.52 & 2.93 \\
    SGMSE+~\cite{richter2023speech}                   & 30 & 2.34 & \textbf{2.95} & \textbf{4.12} & \textbf{3.94} & \textbf{3.62} \\
    ROSE-CD~\cite{xu2025rosecd}                       & 1  & \textbf{2.37} & 2.81 & 4.01 & 3.80 & 3.42 \\
    SBCTM~\cite{nishigori2025schrodinger}             & 1  & 2.24 & 2.78 & 3.83 & 3.88 & 3.33 \\
    MeanFlowSE~\cite{li2026meanflowse}                & 1  & 2.20 & 2.79 & 3.88 & 3.51 & 3.21 \\
    \midrule
    DriftSE (WavLM)                         & 1  & 2.62 & 2.67 & 3.85 & \textbf{3.94} & \textbf{3.42} \\
    DriftSE (HuBERT)                        & 1  & 2.56 & 2.74 & \textbf{3.92} & 3.79 & 3.40 \\
    DriftSE (DistilHuBERT)                 & 1  & \textbf{2.65} & \textbf{2.97} & 3.78 & 3.84 & 3.31 \\
    \midrule    
    DriftSE$^*$ (DistilHuBERT) & 1  & 2.45 & 2.78 & \textbf{4.01} & 3.68 & 3.43 \\
    DriftSE$^\dagger$ (DistilHuBERT) & 1  & \textbf{2.51} & \textbf{2.86} & 4.00 & \textbf{3.82} & \textbf{3.47} \\
    \bottomrule
\end{tabular}}
\vspace{-12pt}
\end{table}

\textbf{Distributional Convergence Visualization:} We qualitatively verify the latent drifting mechanism by tracking a fixed test utterance's evolution in the DistilHuBERT semantic space. Figure~\ref{fig:pca_evolution} reveals a clear transition from a noise distribution toward the clean distribution. While the generated audio initially overlaps with the noisy distribution, optimization of the drifting field enables the model to capture the structural characteristics of the clean distribution. The converged contours and centroids demonstrate that DriftSE successfully maps noisy observations to the high-density regions of the clean speech distribution at equilibrium.

\section{Conclusion}
\label{sec:conclusion}

In this paper, we introduced DriftSE, which achieves native one-step speech enhancement by reformulating denoising as an equilibrium problem. By leveraging a latent drifting field across a multi-scale semantic space, DriftSE successfully aligns its pushforward distribution with the clean speech target, providing a robust signal for high-quality generation. Extensive evaluations confirm that DriftSE achieves state-of-the-art perceptual quality and generalization. Its ability to support fully unpaired learning highlights drifting-based generative modeling as a promising direction with native single-step inference.

\section{Generative AI Use Disclosure}
We acknowledge the ISCA policy stating that generative AI tools cannot serve as co-authors and should only be used for editing or polishing rather than producing significant parts of this paper. Although the proposed method is a novel generative model for speech enhancement, the authors declare that no generative AI tools were used to develop the source code, but AI tools were used to correct text grammar.

\bibliographystyle{IEEEtran}
\bibliography{mybib}
\appendix

\end{document}